\documentclass{IEEEtran}
\usepackage{cite}
\usepackage{amsmath,amssymb,amsfonts}
\usepackage{algorithmic}
\usepackage{graphicx}
\usepackage{textcomp}
\def\BibTeX{{\rm B\kern-.05em{\sc i\kern-.025em b}\kern-.08em
    T\kern-.1667em\lower.7ex\hbox{E}\kern-.125emX}}
\newcommand{\figref}[1]{\figurename~\ref{#1}}

\newcommand{\tabref}[1]{Table~\ref{#1}}
\begin{document}
\title{{\fontsize{24}{26}\selectfont{Communication\rule{29.9pc}{0.5pt}}}\break\fontsize{16}{18}\selectfont
Fast and Automatic 3D Modeling of Antenna Structure Using CNN-LSTM Network for Efficient Data Generation}
\author{Zhaohui~Wei,~\IEEEmembership{Graduate Student Member,~IEEE,}
        Zhao~Zhou,~\IEEEmembership{Graduate Student Member,~IEEE,}
        Peng~Wang,~\IEEEmembership{Graduate Student Member,~IEEE,}
        Jian~Ren,~\IEEEmembership{Member,~IEEE,}
        Yingzeng~Yin,~\IEEEmembership{Member,~IEEE,}
        Gert~Frølund~Pedersen,~\IEEEmembership{Senior Member,~IEEE,}
        and~Ming~Shen,~\IEEEmembership{Senior Member,~IEEE}% <-this % stops a space
\thanks{Manuscript received *** **, ****; revised *** **, ****. This work is sponsored by China Scholarship Council. (Corresponding author: Jian Ren and Ming Shen)}
\thanks{Zhaohui Wei, Zhao Zhou, Peng Wang, Gert Frølund Pedersen, and Ming Shen are with the Department of the Electronic Systems, Aalborg University, 9220 Aalborg, Denmark. (Ming Shen: mish@es.aau.dk).}
\thanks{Jian Ren, and Yingzeng Yin are with the National Key Laboratory of Antennas and Microwave Technology, Xidian University, Xi’an 710071, China. (Jian Ren: renjian@xidian.edu.cn).}
\thanks{Color versions of one or more of the figures in this communication are available online at https://ieeexplore.ieee.org.}
\thanks{Digital Object Identifier: }}
\maketitle

\begin{abstract}
Deep learning-assisted antenna design methods such as surrogate models have gained significant popularity in recent years due to their potential to greatly increase design efficiencies by replacing the time-consuming full-wave electromagnetic (EM) simulations. However, a large number of training data with sufficiently diverse and representative samples (antenna structure parameters, scattering properties, etc.) is mandatory for these methods to ensure good performance. Traditional antenna modeling methods relying on manual model construction and modification are time-consuming and cannot meet the requirement of efficient training data acquisition. In this study, we proposed a deep learning-assisted and image-based intelligent modeling approach for accelerating the data acquisition of antenna samples with different physical structures. Specifically, our method only needs an image of the antenna structure, usually available in scientific publications, as the input while the corresponding modeling codes (VBA language) are generated automatically. The proposed model mainly consists of two parts: Convolutional Neural Network (CNN) and Long Short Term Memory (LSTM) networks. The former is used for capturing features of antenna structure images and the latter is employed to generate the modeling codes. Through training, the proposed model can achieve fast and automatic data acquisition of antenna physical structures based on antenna images. Experiment results show that the proposed method achieves a significant speed enhancement than the manual modeling approach. This approach lays the foundation for efficient data acquisition needed to build robust surrogate models in the future.
\end{abstract}

\IEEEpeerreviewmaketitle

\section{Introduction}
\IEEEPARstart{A}{ntenna} design and optimization is a crucial part of modern communication systems, with applications ranging from wireless networks \cite{wireless} to satellite communications \cite{satellite}. In recent years, machine learning (ML) \cite{classification, regression} has emerged as a powerful tool for automating the antenna design process, with the potential to improve efficiency and accuracy. One of the most popular methods for using ML in antenna design is through the creation of surrogate models \cite{surrogate1,surrogate2,surrogate3,surrogate4,surrogate5,surrogate6,surrogate7}, which fully replace the computationally expensive EM simulations. In this process, the surrogate models mainly work by mapping the design variables of physical structures to the EM responses. The trained models can accurately and efficiently predict the performance of new antenna designs based on a limited set of training data. Many surrogate model-aided numerical optimization methods have been proven successful in the antenna design area. In \cite{surrogate3, surrogate7}, the surrogate models were built to quickly predict the electrical and radiative properties of a planar ultra-wideband dipole antenna and microstrip antenna subarrays, respectively. In \cite{surrogate4}, the author proposed a hybrid surrogate model-assisted optimization framework for solving complex antenna design with 45 design variables and 18 specifications. These methods show good performance in the efficiency enhancement of antenna design, but the surrogate model in these methods can only predict the physical properties of a single type of antenna. To improve the capabilities of the surrogate model, a diversity of EM structures is used as training data. In \cite{generative1, geneartive2, generative3}, multiple basic meta-units are employed to generate different kinds of metasurfaces or selective frequency surfaces for training the surrogate model. As a result, the method can select the feasible solution from multiple structures and have higher probabilities to generate the optimal result.

\begin{figure}[t!] %!t
\centering
\includegraphics[width=3.4in]{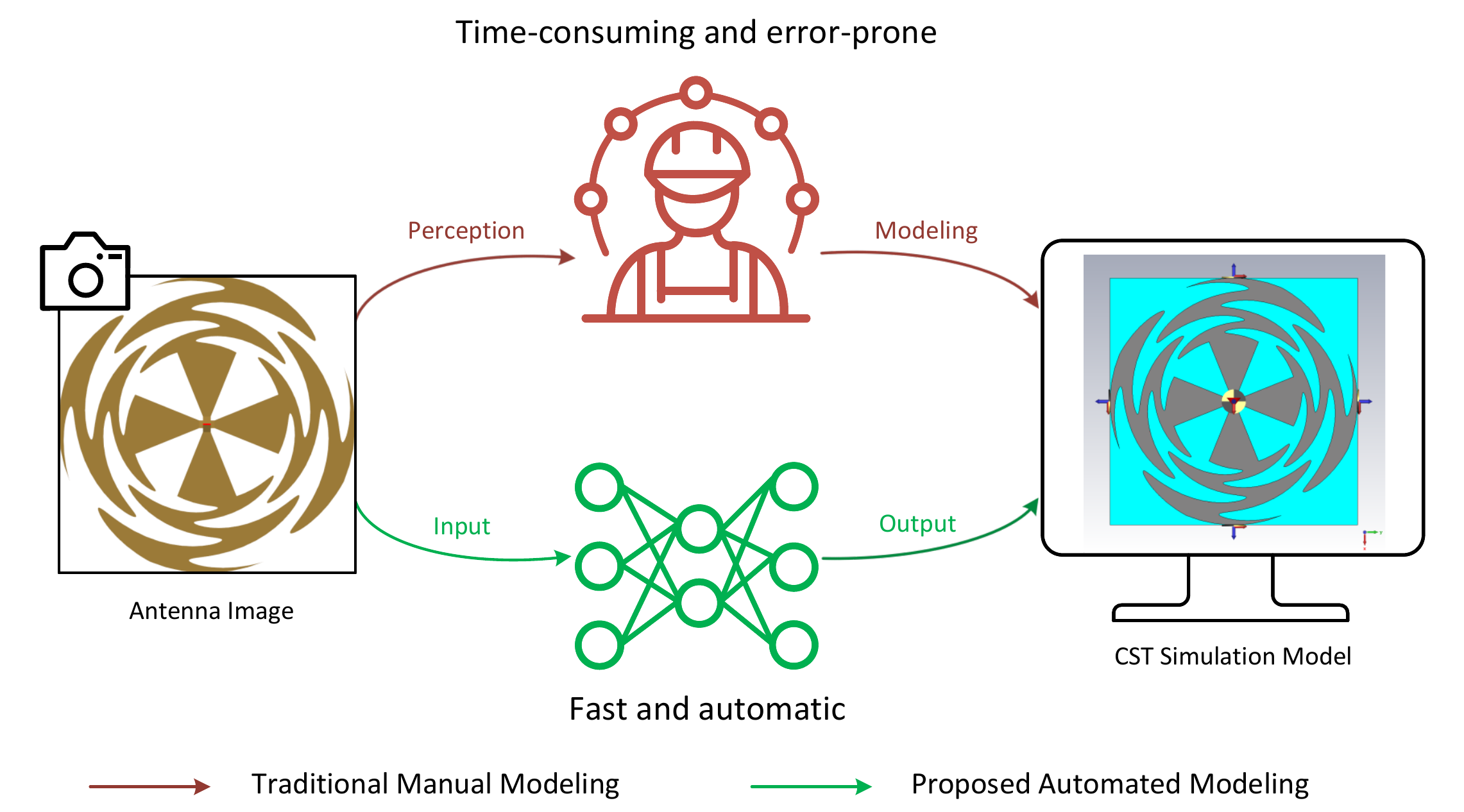}
\caption{Comparison of traditional manual antenna modeling (top) and proposed automated modeling (bottom). Manual modeling involves significant time and effort from antenna experts, while automated modeling streamlines the process.} 
\label{Manual modeling}
\end{figure}

\begin{figure*}[t!] %!t
\centering
\includegraphics[width=6.6in]{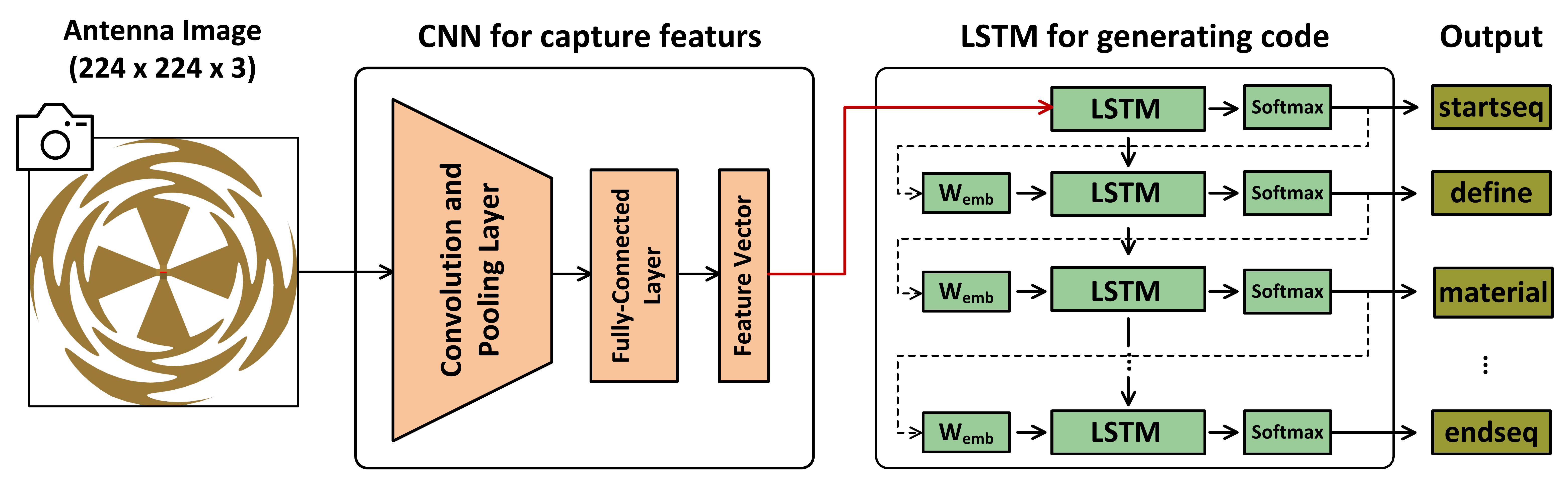}
\caption{Overview of the CNN-LSTM architecture for modeling code generation from antenna images.} 
\label{proposed method}
\end{figure*}
While it is an effective method to boost the capability of the surrogate model by increasing the diversity of training data, the efficient acquisition of training data is a great challenge. As shown in \figref{Manual modeling},  designers usually obtain the training data by creating a 3D model of an antenna using EM simulation software such as CST Microwave Studio. This process involves constructing a precise representation of the antenna's geometrical details, such as its shape, size, and material properties, and any additional components such as feedlines or matching networks, and can be a time-consuming and error-prone process, particularly for complex antenna designs. 

In this study, we propose a fast and automatic modeling method to improve the efficiency of data acquisition. This is done by combining CNN and LSTM networks to automatically generate antenna modeling codes. This approach can significantly reduce the time and effort required for manual modeling, which is crucial for accelerating the design process and increasing the efficiency of deep learning-based antenna design.

\section{Problem Analysis and Formulation}
In contrast to the traditional manual modeling approach, which involves direct human interaction with EM simulation software, the automated modeling approach directly generates modeling code to control the EM software and create the antenna model. Therefore, the automated antenna modeling problem can be converted into a modeling code generation problem. In our study, we focus on the generation of Visual Basic for Application (VBA) as CST uses the VBA language for modeling 3D structures. 

The problem of modeling code generation from antenna images can be formulated as a regression problem, which is to find the mapping between the input space of antenna images and the output space of modeling code sequences. The input space $X$ consists of $N$ antenna images, each represented by a matrix $x_i$ of size $m \times m$. The output space $Y$ consists of $N$ modeling code sequences, each represented by a vector $y_i$ of length $l_i$. The mapping function $f:X \rightarrow Y$ is used to generate a modeling code sequence $y_i$ for a given antenna image $x_i$. The objective is to learn this mapping function from a set of training examples $(x_1, y_1), (x_2, y_2), ..., (x_n, y_n)$ such that it can accurately predict the CST modeling code sequence for new, unseen antenna images.

To solve this problem, we choose a machine-learning model to approximate the mapping function. When selecting the machine learning model for generating CST modeling code from antenna images, the input data and output data characteristics should be taken into consideration. Antenna images are spatially represented as 2D matrices that can be best modeled using  CNNs capable of effectively extracting spatial features and learning hierarchical representations. Conversely, CST modeling code sequences consist of a series of commands describing antenna geometry and material properties, making recurrent neural networks (RNNs) such as LSTM networks an ideal choice as they can capture temporal dependencies in the sequence data and generate output sequences of varying lengths.

To combine these models for joint learning, a CNN-LSTM architecture is typically used. In this approach, CNN is used for feature extraction from the input data, while LSTM generates the output sequence based on the extracted features. The CNN-LSTM model has demonstrated effectiveness in various sequence-to-sequence tasks and is a suitable model for generating CST modeling code from antenna images.

\section{The Proposed Method}
\subsection{Overview of the CNN-LSTM Architecture}
As shown in \figref{proposed method}, a graphical representation of the automated process for modeling code generation using antenna images is illustrated. The proposed method employs a CNN-LSTM architecture to generate modeling code for CST simulations using antenna images as input. The CNN component is responsible for extracting relevant features from the input images, while the LSTM component generates sequences of modeling code based on these features. The model is trained on a dataset of antenna images and corresponding modeling code to learn the patterns and relationships between them. In the inference phase, the model takes an antenna image as input and generates modeling code that can be used to reproduce the antenna design in CST. This method has the potential to significantly simplify the process of creating antenna models for CST simulations, offering an efficient and accessible solution for users. The CNN and LSTM components used in the proposed method will be outlined in detail in the following subsections.

\begin{figure}[t!] %!t
\centering
\includegraphics[width=3.4in]{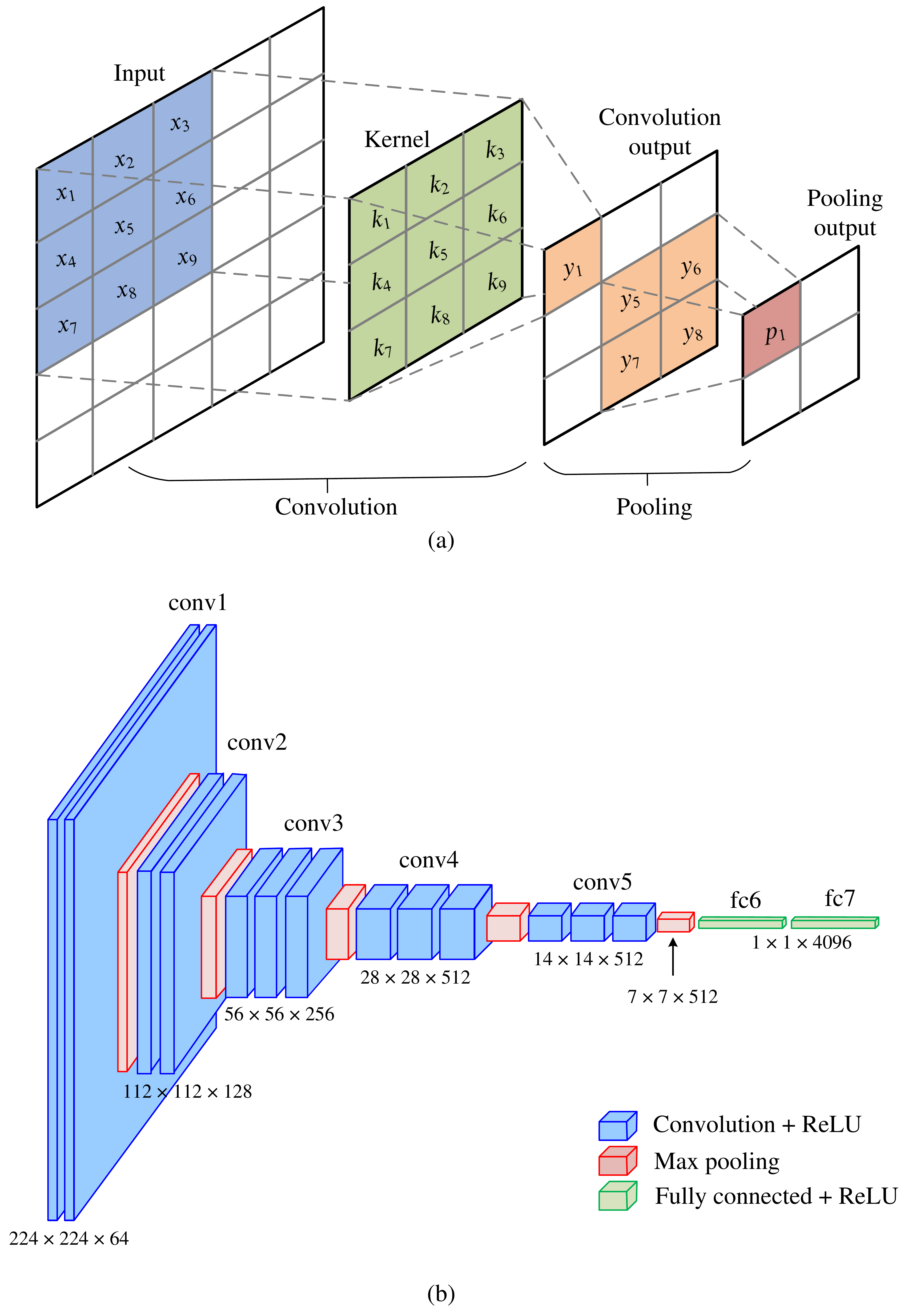}
\caption{(a) Convolution and pooling operations applied to a 2D input image. (b) Modified VGG16 model used in our study for image features extraction.} 
\label{CNN}
\end{figure}

\subsection{CNN-based Feature Extractor}
Convolution and pooling are two fundamental operations in CNN that enable the network to learn hierarchical representations of input data, such as images or audio. As shown in \figref{CNN}(a), convolution is a linear operation that involves sliding a kernel (a small matrix of numerical coefficients) over an input signal or image and computing the sum of the element-wise product of the kernel and the input at each position. The resulting output is a set of feature maps that capture different aspects of the input data. In a CNN, the convolution operation applies a set of learnable filters (kernels) to the input image, resulting in a set of feature maps that capture different aspects of the input image. Each filter responds to a different feature or pattern in the input, such as edges, corners, or textures. By stacking multiple convolutional layers on top of each other, the network can learn increasingly complex and abstract features of the input data. Pooling is a non-linear operation that involves down-sampling the feature maps obtained from the convolutional layers. The most common form of pooling is max pooling, which involves dividing the input feature map into non-overlapping sub-regions and taking the maximum value within each sub-region. The resulting output is a set of pooled feature maps that are smaller in size but retain the most salient features of the input. Pooling helps to reduce the spatial dimensionality of the feature maps, making the network more computationally efficient and reducing the risk of over-fitting.

In our proposed method, we utilized a modified version of the VGG16 model as the CNN component. The VGG16 model \cite{VGG16} is a widely used CNN architecture that was introduced by the Visual Geometry Group (VGG) at the University of Oxford in 2014. As shown in \figref{CNN}(b), the model is based on a sequential arrangement of convolutional layers, with five groups of convolutional layers followed by two fully connected layers. The first two convolutional groups consist of two convolutional layers each, while the remaining three groups each have three convolutional layers. The convolutional layers in the VGG16 model use 3$\times$3 filters with a stride of 1 and the same padding to maintain the spatial dimensions of the input. The activation function used in the model is Rectified Linear Units (ReLU), which has been shown to provide improved performance compared to traditional activation functions such as sigmoid or tanh. The pooling layers in the VGG16 model use 2$\times$2 windows with a stride of 2, resulting in a halving of the spatial dimensions of the feature maps. During training, the model is optimized using the categorical cross-entropy loss function and the Adam algorithm. The weights of the model are initialized using the pre-trained weights of the VGG16 model on the ImageNet dataset \cite{Imagenet}, which is a large-scale image recognition dataset containing millions of images across 1,000 categories. After training, the CNN model can extract high-level features from the input antenna images, which are used to feed the following LSTM model for generating the corresponding modeling codes. 

\begin{figure}[t!] %!t
\centering
\includegraphics[width=2.8in]{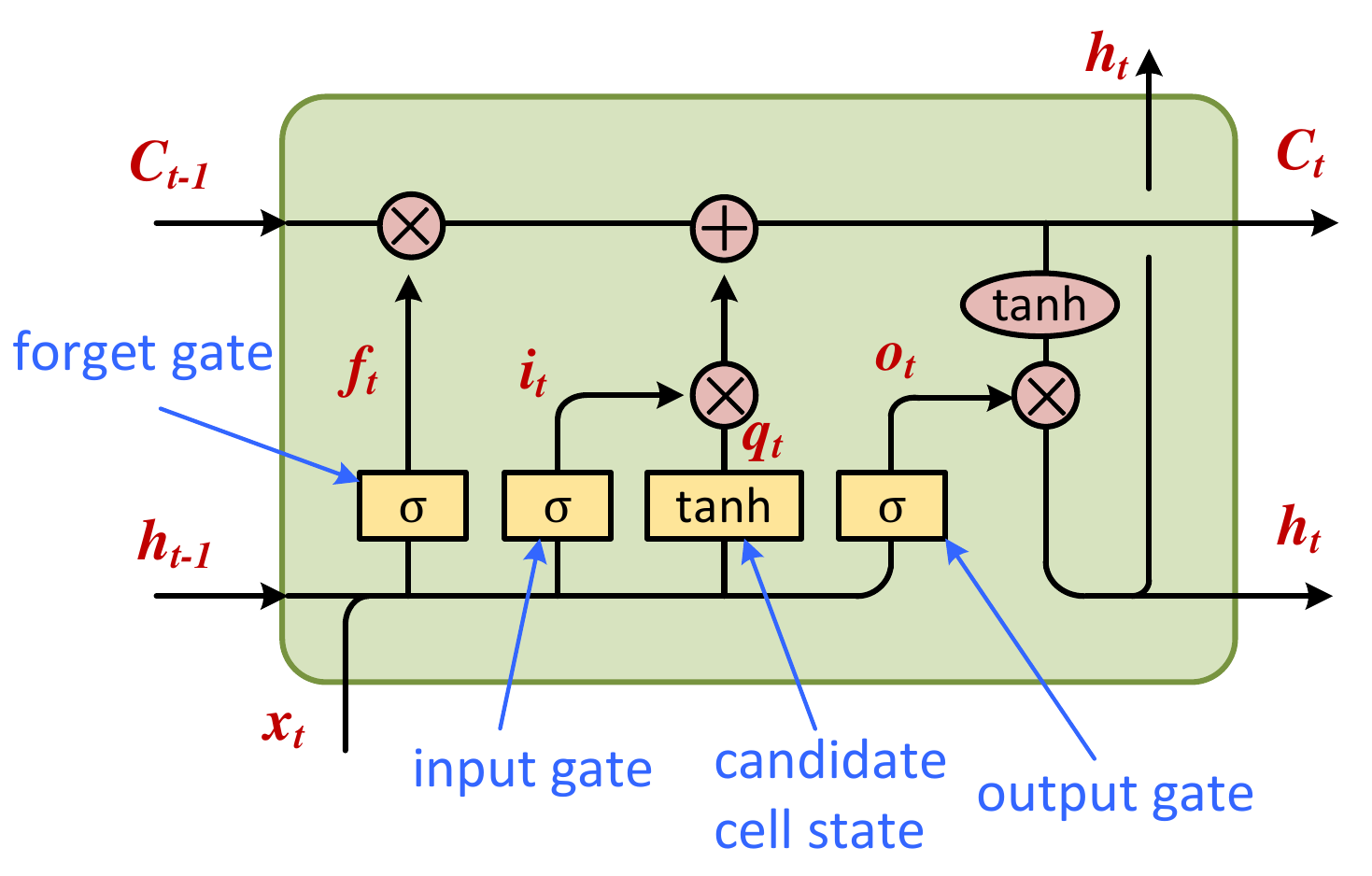}
\caption{Schematic representation of a single LSTM cell.} 
\label{LSTM}
\end{figure}

\subsection{LSTM-based Code Generator}
LSTM is a type of recurrent neural network that has been widely used in various machine learning applications, especially in processing sequential data. The LSTM architecture \cite{LSTM} was first introduced by Hochreiter and Schmidhuber in 1997 as a solution to the vanishing gradient problem in traditional recurrent neural networks. In recent years, LSTM has become one of the most popular and powerful tools in the field of deep learning, with numerous successful applications across a wide range of domains. Its ability to handle long-term dependencies and process sequential data has made it a critical component in many cutting-edge machine learning systems.

\figref{LSTM} shows the structure of LSTM. LSTM units contain several interconnected neural networks that control the flow of information through the network. The input gate determines which new information to store in the cell state, while the forget gate decides which information to discard from the previous cell state. The cell state stores long-term information that can be modified or retrieved as needed, and the output gate regulates how much information to reveal to the next hidden state. The gating mechanisms in LSTM units allow them to selectively process and filter information, making them especially useful for tasks that require processing and analyzing long sequences of data. The detailed data flow can be represented using the following formula.

\begin{figure}[t!] %!t
\centering
\includegraphics[width=3.2in]{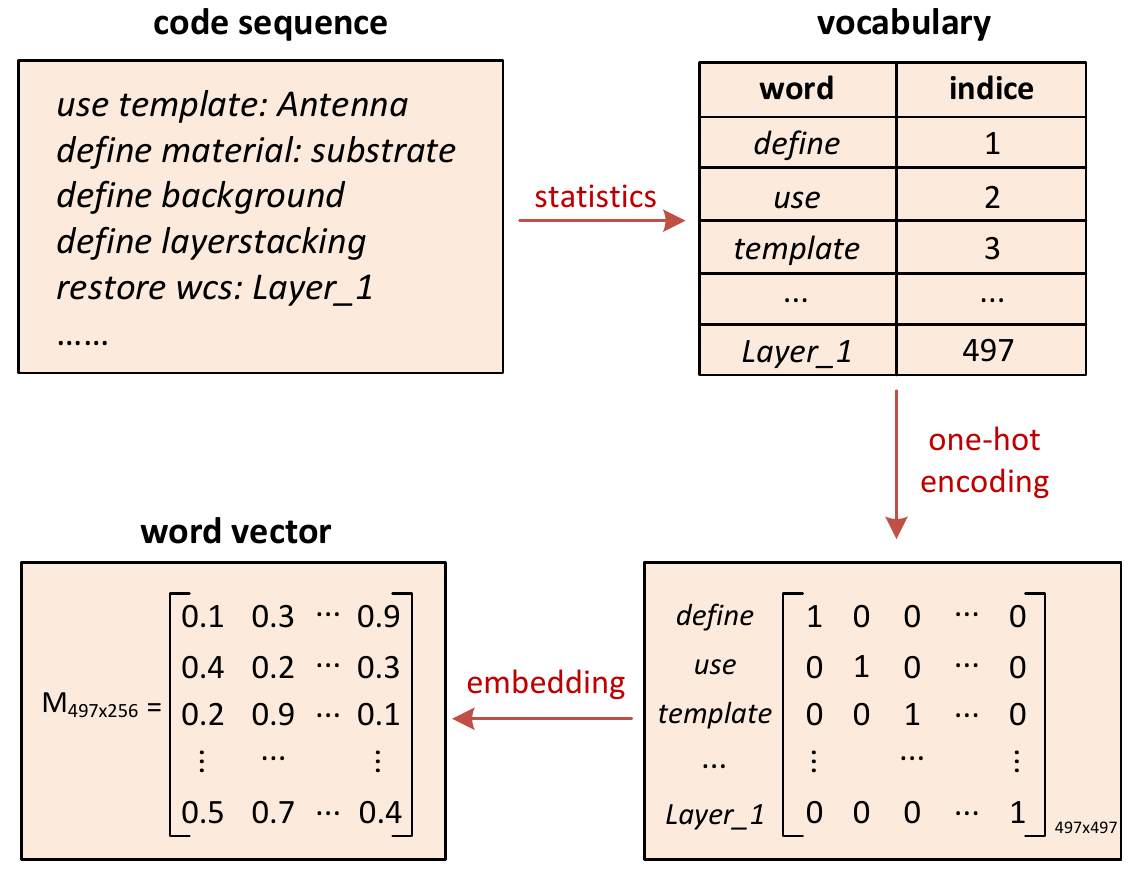}
\caption{ A diagram of transforming code sequences into word vectors.} 
\label{embedding}
\end{figure}

\begin{equation}
f_t=\sigma\left(W_f \cdot [x_t,h_{t-1}]+b_f \right),
\end{equation}
\begin{equation}
i_t=\sigma\left(W_i \cdot [x_t,h_{t-1}] + b_i\right),
\end{equation}
\begin{equation}
q_t=\tanh \left(W_q \cdot [x_t,h_{t-1}]+b_q\right),
\end{equation}
\begin{equation}
C_t=f_t * C_{t-1}+i_{t} * q_{t},
\end{equation}
\begin{equation}
o_t=\sigma\left(W_o \cdot [x_t,h_{t-1}]+b_o\right),
\end{equation}
\begin{equation}
h_t=o_t * \tanh(C_t),
\end{equation}

\noindent where,
$f_t$ is the output of the forget gate at time step $t$,
$i_t$ is the output of the input gate at time step $t$,
$q_t$ is the candidate cell state at time step $t$,
$C_t$ is the current cell state at time step $t$,
$o_t$ is the output of the output gate at time step $t$,
$h_t$ is the current hidden state at time step $t$,
$\sigma$ is the sigmoid activation function,
$\tanh$ is the hyperbolic tangent activation function,
$W$ and $U$ are weight matrices for the input $x_t$ and previous hidden state $h_{t-1}$, respectively,
$b$ is the bias term.

\begin{table}[t!]
\caption{An Example of Converting Code Sequence Into Input and Output for Training LSTM}
\begin{center}
\small
\begin{tabular}{ | c | c |}
\hline
\rule{0pt}{11pt} %改变行高
 Input (code sequence) & Output (code to predict) \\ \hline
\rule{0pt}{11pt} %改变行高
\textit{use} & \textit{template}\\\hline
\rule{0pt}{11pt} %改变行高
\textit{use}, \textit{template} & \textit{Antenna} \\\hline
\rule{0pt}{11pt} %改变行高
\textit{use}, \textit{template}, \textit{Antenna} & \textit{define} \\\hline
\rule{0pt}{11pt} %改变行高
\textit{use}, \textit{template}, \textit{Antenna}, ... & ... \\\hline
\end{tabular}
\label{1}
\end{center}
\end{table}
One important preprocessing step for training an LSTM model is to transform the code sequence into word vectors. \figref{embedding} illustrates the overall process of transforming code sequences into word vectors. First, the vocabulary of the code sequence is constructed by collecting all unique words that appear in the code sequence. This vocabulary serves as the basis for the subsequent vectorization process. Next, each word in the code sequence is represented as a one-hot vector, where the vector has the same dimension as the vocabulary size and contains a 1 at the index corresponding to the word's position in the vocabulary and 0s elsewhere. This representation ensures that each word in the sequence is uniquely identified by its position in the vocabulary, thereby enabling the LSTM model to learn the statistical patterns and relationships between the words.
However, the one-hot representation suffers from high dimensionality and sparsity, which can be problematic for training deep neural networks. Therefore, an embedding operation is applied to transform the one-hot vectors into dense and continuous word vectors, which capture the semantic and syntactic properties of the words in the sequence. This operation involves mapping each one-hot vector to a low-dimensional vector space (typically in the range of 50-300 dimensions, 256 is selected in this study) using a trainable weight matrix. After embedding, each word in the code sequence can be converted into a unique vector which can be calculated by computers. 

Modeling code sequence generation is a supervised learning task that requires the provision of input and output pairs for training an LSTM model. To accomplish this, a common approach is to employ a sliding window technique that divides the code sequence into fixed-length subsequences, with each subsequence serving as an input to the LSTM model. The output associated with each input is determined by the following word in the sequence. \tabref{1} gives an example of converting code sequence into input and output. Consider the code sequence ``[\textit{use template Antenna}, \textit{define material},...]". In this case, the first input to the LSTM model would be the word ``\textit{use}", with the corresponding label being the subsequent word ``\textit{template}". The second input would then be the words ``\textit{use template}", with the corresponding label being ``\textit{define}". This process is repeated until the end of the sequence is reached, with each subsequence contributing to the formation of multiple input-output pairs. 

\section{Experiments and Results}

\subsection{Data Collection and Preprocessing}
\begin{figure}[t!] %!t
\centering
\includegraphics[width=3.2in]{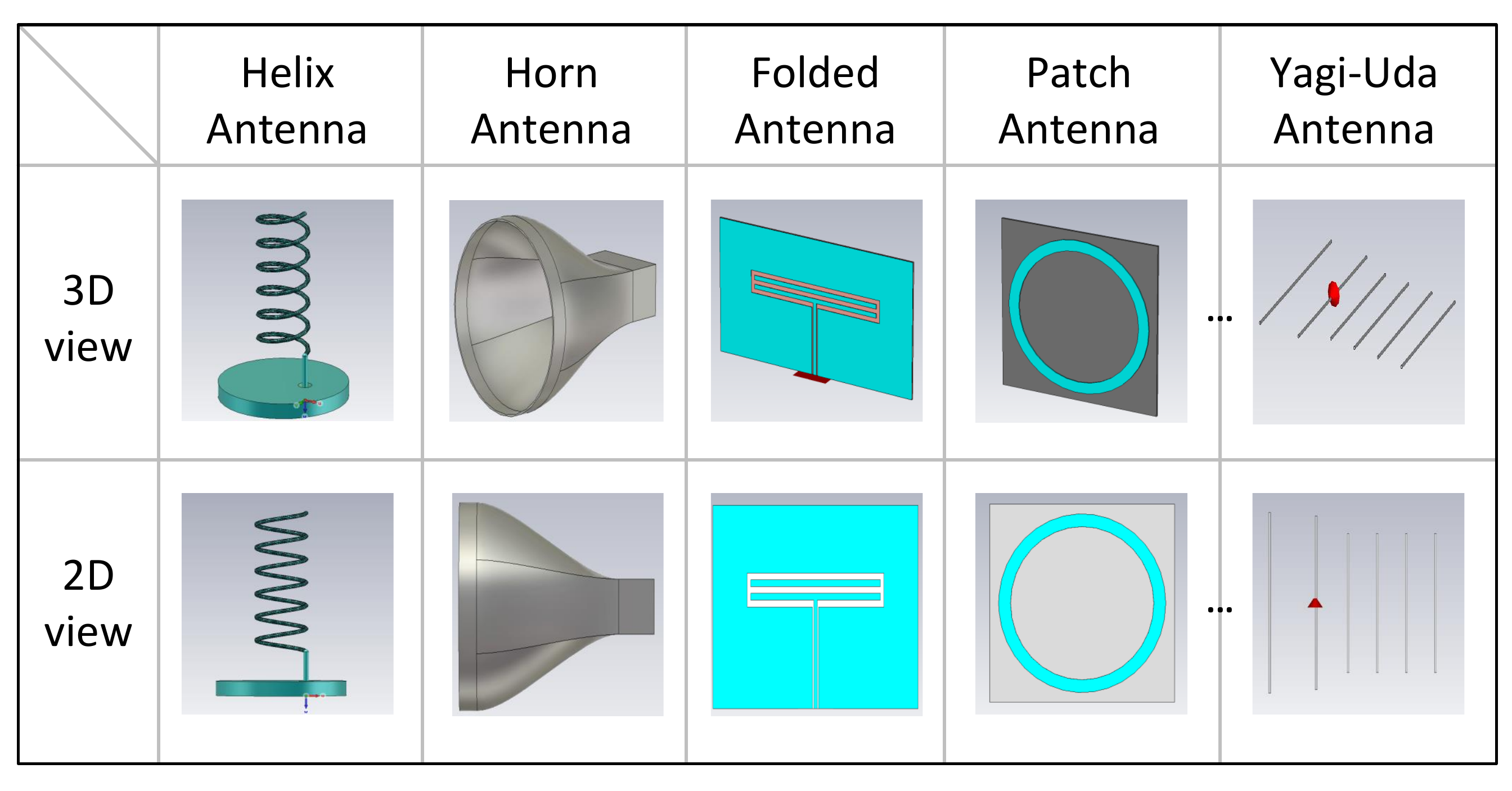}
\caption{Antenna image dataset for developing the CNN-LSTM model for automating modeling code generation.} 
\label{images}
\end{figure}

\figref{images} shows the antenna image dataset for developing the CNN-LSTM model for automating modeling code generation. In our study, a set of 200 antenna images covering about 20 different antenna categories and their corresponding modeling codes are collected. Considering the accuracy and efficiency of antenna modeling, each antenna image is collected from two views (a 3D view and a 2D view) for providing geometrical information. The collection of both the antenna images and their corresponding modeling codes was performed using CST.

Before training the machine learning model, the image data needs to be preprocessed, which mainly involves two steps. The input image by first resizing the image to a fixed size of 224 $\times$ 224 pixels to ensure that all input images have the same size. After resizing, the color antenna images need to be converted into grayscale images for removing the distracting features.

\subsection{Training Details}
\begin{figure}[t!] %!t
\centering
\includegraphics[width=3.2in]{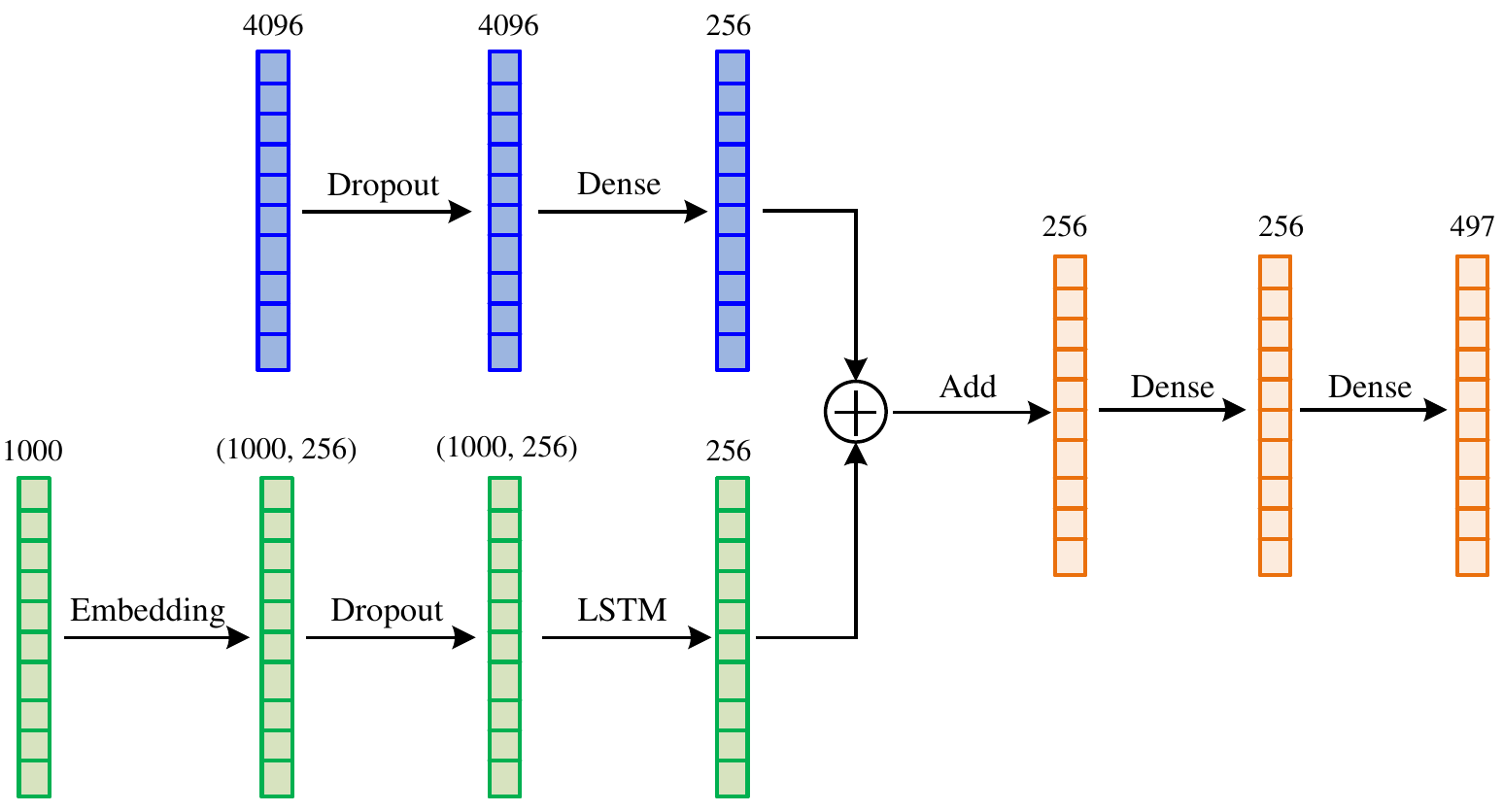}
\caption{Data flow of the proposed method.} 
\label{training}
\end{figure}

The dataset was partitioned into training, validation, and testing subsets, comprising 70$\%$, 15$\%$, and 15$\%$ of the data, respectively. To extract meaningful features from input antenna images, the CNN component of the model was pre-trained on the ImageNet dataset, while the LSTM component was trained to generate modeling code based on these features. A schematic of the proposed model's data flow is depicted in \figref{training}, wherein the output of the CNN feature extractor comprises 4096-dimensional vectors, which are subsequently processed by a fully connected layer to produce a 256-element representation of the antenna images. Each code sequence, with a maximum length of 1000, is transformed into a 1000$\times$256 matrix using the embedding layer and fed into an LSTM layer with 256 recurrent units. The 256-dimensional image and code sequence representations are fused using an additional operation and then passed through a Dense layer with 256 neurons, followed by the final output Dense layer, which generates softmax predictions for the next word in the sequence across all output code vocabulary. Additionally, a 50$\%$ dropout rate is employed to mitigate overfitting to the training dataset.

\begin{figure}[t!] %!t
\centering
\includegraphics[width=3.4in]{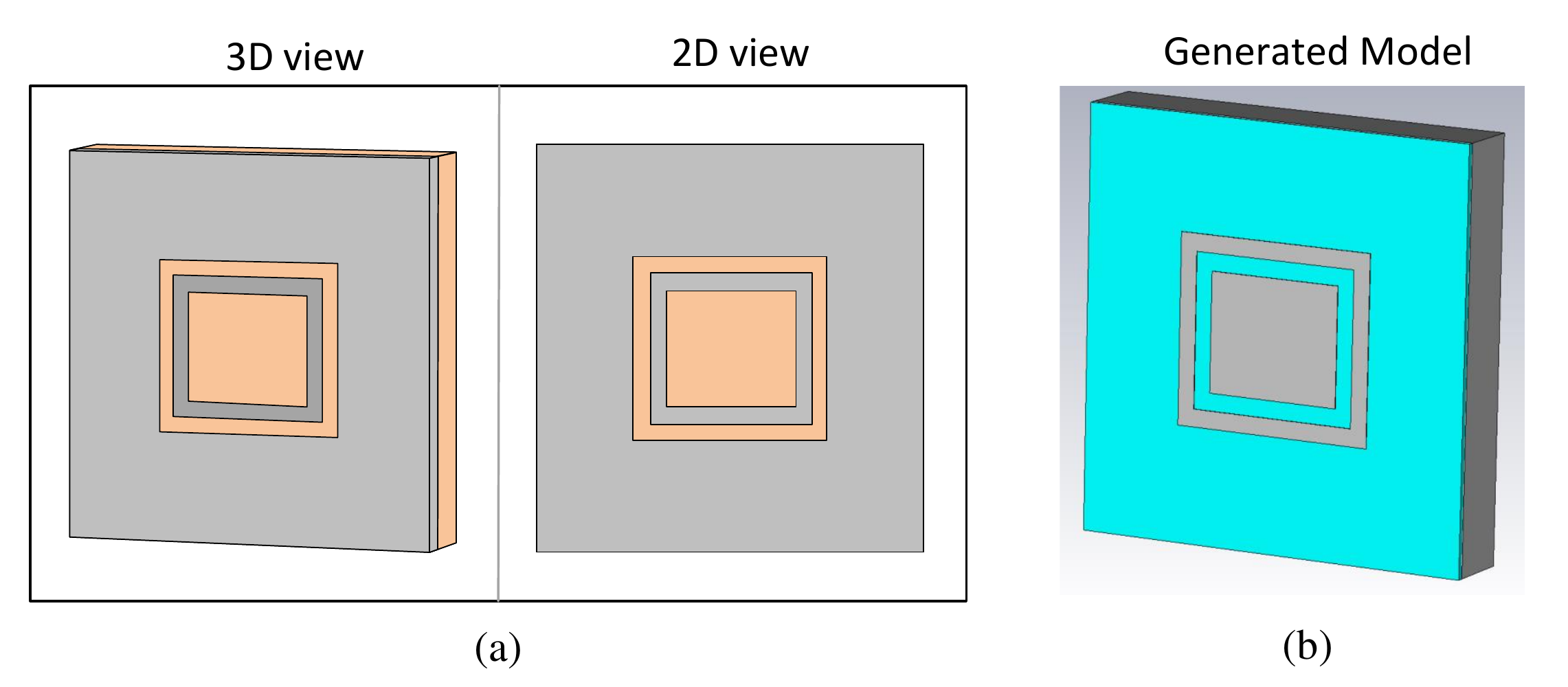}
\caption{Testing results. (a) Input antenna images. (b) Generated antenna physical model.} 
\label{Test}
\end{figure}

During the training process, we used the categorical cross-entropy loss function to minimize the difference between the predicted modeling code and the ground truth modeling code. We also utilized the Adam optimizer to update the parameters of the model during training. The model was trained for 50 epochs with each epoch taking approximately 307 seconds to complete. The training loss decreases with each epoch and reaches a low value of 0.005 after 50 epochs, indicating that the model is fitting the training data well. After training, the model was evaluated on the testing set to measure its performance in generating accurate modeling code from antenna images. \figref{Test} gives the test images and the generated antenna physical model. Results show that the trained model can successfully generate the correct modeling code. The corresponding modeling code is placed in the support file. 

\section{Comparison and Discussion}
\subsection{Comparison}
To evaluate the efficiency of the proposed CNN-LSTM-based automatic modeling approach for data acquisition, we conducted a comparative study with the traditional manual modeling approach. The study involved five antenna designers of different levels of expertise, and three images from the antenna test set were modeled using both methods. The modeling times for the manual approach were found to vary between a few minutes to ten minutes, depending on the modeling level of the designer. In contrast, the automatic modeling approach required an average time of around 50 seconds, with the processing time being dependent on the running speed of the neural network and the hardware facilities employed in the study.  Faster computing resources can enable more rapid and efficient modeling.

The findings of this study highlight the advantages of the CNN-LSTM-based automatic modeling approach over the traditional manual approach in terms of efficiency and consistency. The significant variation in modeling times observed for the manual approach, depending on the expertise level of the designer, indicates the potential for inconsistencies in results and the need for significant time investment. In contrast, the automatic modeling approach offers consistent and reliable results within a short time frame, regardless of the level of expertise of the user.

\subsection{Discussion}
While the proposed approach shows great potential for accelerating the physical modeling of antennas, there are still some weaknesses and limitations that need to be addressed to make this method more complete. One of the main weaknesses is that the proposed method relies heavily on the quality and quantity of the training dataset. The accuracy and generalization ability of the model depends on the diversity and completeness of the training dataset. However, collecting a large and representative dataset of antenna images can be time-consuming and challenging, as it requires significant effort and resources to obtain high-quality images that cover various types of antennas, configurations, and materials. 

\section{Conclusion}
In this study, we propose a deep learning-assisted and image-based modeling approach for accelerating the data acquisition of antenna design samples with different physical structures. The method overcomes the limitations of manual modeling approaches and provides a fast and automatic modeling solution based on antenna images. The combination of CNN and LSTM networks enables the model to capture the features of antenna structure images and generate the corresponding EM model codes. Experiment results show that the proposed method achieves a considerable times speed enhancement compared to traditional manual modeling approaches. The proposed approach can significantly reduce the amount of time needed to build robust surrogate models, which is a critical advantage for efficient training data acquisition. This study contributes to the development of deep learning-assisted antenna design and provides a new perspective on accelerating the physical modeling of antennas.
\ifCLASSOPTIONcaptionsoff
  \newpage
\fi

\bibliographystyle{IEEEtran}
\bibliography{IEEEabrv,ref}

\vfill
\end{document}